\documentclass[a4paper,oneside,12pt,english]{article}
\usepackage[pdftex]{color,graphicx}

\usepackage[T1]{fontenc}
\usepackage[latin9]{inputenc}
\usepackage{url}

\usepackage{amsmath}
\usepackage{amssymb}

\makeatletter
\usepackage[backref=page]{hyperref}
\hypersetup{plainpages = false,
              breaklinks=true,
              a4paper=true,
              linktocpage,
              pdfauthor={Niko Brummer},
              pdfstartpage=1, 
              pdfstartview=FitH,  
              pdfkeywords={}}

\makeatother

\usepackage{babel}

\def\Amat{\mathbf{A}}

\def\xvec{\mathbf{x}}

\def\svec{\mathbf{s}}

\DeclareMathOperator{\BD}{Beta}

\begin{document}

\title{Tutorial for Bayesian forensic likelihood ratio}
\date{October 2011}

\author{Niko Br\"{u}mmer\\AGNITIO Research, South Africa}

\maketitle

\section{Introduction}
In the Bayesian paradigm for presenting forensic evidence to court, it is recommended that the weight of the evidence be summarized as a \emph{likelihood ratio} (LR) between two opposing hypotheses of how the evidence could have been produced. Such LRs are necessarily based on probabilistic models, the parameters of which may be uncertain. It has been suggested by some authors that the value of the LR, being a function of the model parameters should therefore also be considered uncertain and that this uncertainty should be communicated to the court.

In this tutorial, we consider a simple example of a \emph{fully Bayesian} solution, where model uncertainty is integrated out to produce a value for the LR which is \emph{not} uncertain. We show that this solution agrees with common sense. In particular, the LR magnitude is a function of the amount of data that is available to estimate the model parameters.

Bayesian methods are often criticised because of the difficulty of choosing appropriate priors, especially when the priors are non-informative. We do not deny these difficulties, but the problem is not solved by adopting frequentist methods that effectively sweep the prior under the carpet and pretend it does not exist. In this tutorial we do need to choose a non-informative prior and we choose it by examining the effect it has on the end-result.  

We shall reference the following books: E.T.~Jaynes, \emph{Probability Theory: The Logic of Science}, Cambridge University Press 2003, which we shall abbreviate as PTLOS; and D.J.~Balding, Weight-of-evidence for Forensic DNA Profiles, Wiley 2005, abbreviated as WEFDNA.

\section{Simplified DNA model}
\def\pvec{\mathbf{p}}
\def\nvec{\mathbf{n}}
\def\xvec{\mathbf{x}}
\def\qvec{\mathbf{q}}
\def\avec{\mathbf{a}}
In this tutorial we shall derive the details of how to compute the LR with a \emph{simplified} DNA-like model. The idea is not to provide a recipe that can be used in real forensic DNA analysis, but rather to choose a model that facilitates better understanding of  the basic look and feel of a fully Bayesian solution. We need the model to be very simple so that we can perform the Bayesian integrals in closed form. More realistic models would require more complex methods, which would obscure the primary purpose of this tutorial.

We suppose that the \emph{DNA profile} of every individual has $K$ different binary \emph{loci} the \emph{state} of each of which can be either 1 or 0.  Every individual is therefore categorized by $K$ binary variables, which gives a total number of $2^K$ states.\footnote{In real DNA profiling, there are different locus types, with more complex state spaces. For example, STR loci consist of two parts with independent states, one inherited from the father and the other from the mother. Each part has 2 or more states, called \emph{alleles}. DNA profiling technology can detect the state of each part, but does not show which comes from the mother and which from the father.} We represent a DNA profile by a vector of the form $\avec=(a_1,a_2,\ldots,a_K)$, where $a_k\in\{0,1\}$ represents the state of locus $k$.

We assume that given a DNA sample (either recovered at the crime scene where it was left by the \emph{perpetrator}, or obtained from the \emph{suspect}), the state of each locus may be determined without error.

The main complication is when all suspect and perpetrator loci match, that there is a non-zero probability that some person other than the suspect could have the same DNA profile. To compute this probability, we need to model profile distributions.

\section{Profile distribution model}
Here we define a generative model that is probably about as simple as it can be. Again, our goal is just to illustrate the basic principles of a fully Bayesian approach to this kind of problem. The goal of this exercise is not to reproduce a realistic DNA model---in real population genetics, the models are more complex.

Let the probability that locus $k$ of a randomly chosen person has state 1 be $q_k$, and the probability that it has state 0 be $1-q_k$. According to this model we assume the following independencies:
\begin{itemize}
	\item The locus states are independent: knowing the state of locus $k$ for one or more individuals, tells us nothing about the states of other loci $k'$.
	\item For each locus $k$, the binary state for each person is sampled as an \emph{iid} Bernoulli trial with parameter $q_k$.  
\end{itemize}
We can collect the locus probabilities in the vector\footnote{Note that the elements of $\qvec$ usually do not sum to one. These are $K$ independent probabilities, not one $K$-ary categorical distribution.} $\qvec=(q_1,q_2,\ldots,q_K)$. We refer to $\qvec$ as the \emph{model parameter}, which encodes everything there is to know (under the above modelling assumptions) about how locus states are distributed in the population. The model can be summarized by:
\begin{align}
P(\avec|\qvec) &= \prod_{k=1}^K q_k^{a_k}(1-q_k)^{1-a_k}
\end{align}
which is the probability that a randomly chosen individual has DNA profile $\avec=(a_1,a_2,\ldots,a_K)$ in a population characterized by the model parameter $\qvec=(q_1,q_2,\ldots,q_K)$. The complication is that we are \emph{not} given $\qvec$. Its value has to be inferred from prior assumptions and from data.

\section{Inferring the model parameter}
\def\pivec{\boldsymbol{\pi}}
We do a Bayesian inference for the value of $\qvec$, by computing a posterior distribution. 

\subsection{Prior}
As prior for $q_k$, we assign a \emph{beta distribution}. This choice has a threefold motivation: (i) The beta distribution is a conjugate prior for this problem, which allows for closed-form Bayesian calculations. (ii) It is commonly used in forensic DNA practice. (iii) It is general enough to include various non-informative priors, which will be of special interest to us.

We assign independently for each $q_k$ a beta distribution with hyper-parameter $\pi_k = (\alpha_k,\beta_k)$, so that:
\begin{align}
P(\qvec|\pivec) &= \prod_{k=1}^K \BD(q_k|\alpha_k,\beta_k) \\
&= \prod_{k=1}^K \frac{q_k^{\alpha_k-1}(1-q_k)^{\beta_k-1}}{B(\alpha_k,\beta_k)}
\end{align}
where we have defined $\pivec=(\pi_1,\pi_2,\ldots,\pi_K)$. The normalization constant of the beta distribution is given by the \emph{beta function}, defined as:
\begin{align}
B(\alpha,\beta)&=\frac{\Gamma(\alpha)\Gamma(\beta)}{\Gamma(\alpha+\beta)} 
= \int_0^1 q^{\alpha-1}(1-q)^{\beta-1} \,dq
\end{align}
where $\Gamma$ is the gamma function. 

For the beta distribution to be normalized, we need $\alpha_k,\beta_k>0$ and unless stated otherwise, we shall assume this condition holds for all our calculations below. In places, we will however consider the limit as $\alpha_k=\beta_k\to0$. When we do this, we will follow the advice of PTLOS and complete the whole calculation under the assumption $\alpha_k,\beta_k>0$ and apply the limit only to the final result.

\subsubsection{Non-informative priors}
\label{sec:non}
If we want to use a non-informative prior, we let $\alpha=\alpha_k=\beta_k$ by symmetry, and we can choose some $\alpha$, for example in the range $0<\alpha\le1$. The case $\alpha\to0$ is called the \emph{Haldane} prior, the case $\alpha=0.5$ is the \emph{Jeffreys} prior and $\alpha=1$ is the \emph{Laplace} prior.

The Haldane prior is \emph{flat} in the sense that the probability density for $\log\frac{q}{1-q}$ is uniform, but since this reparametrization of $q$ covers the whole real line, this prior is improper.

The Jeffreys prior is flat in the sense that the probability density for $\arcsin(2q-1)$ is uniform between $-\frac{\pi}{2}$ and $\frac{\pi}{2}$. 

The Laplace prior is flat in the sense that the probability density for $q$ is uniform between 0 and 1. 

As these names show, different workers in probability theory have arrived at different conclusions about which prior should be used to encode non-informativeness about the Bernoulli model parameter. To make our calculations concrete, we will have to make a definite choice of prior. We shall solve this problem in a later section, by examining the effect of the prior on the end-result of our calculation.

\subsubsection{Informative prior}
In forensic DNA\footnote{See WEFDNA pp. 63-64.} it is customary to reparametrize the beta prior as:
\begin{align}
\alpha_k &= \frac{1-\theta}{\theta}p_k\,, & \beta_k &= \frac{1-\theta}{\theta}(1-p_k) 
\end{align}
where $0<p_k<1$ and $0<\theta<1$. Here $\theta$ is known as the \emph{population structure parameter}. With this parametrization, $\BD(q_k|\alpha_k,\beta_k)$ has the following mean and variance:
\begin{align}
\langle q_k\rangle &= \frac{\alpha_k}{\alpha_k+\beta_k} = p_k & \langle(q_k-p_k)^2\rangle &= \theta p_k(1-p_k)
\end{align}
For small values of $\theta$, one obtains an \emph{informative} prior, with a small variance and a sharp peak near $p_k$. In the extreme as $\theta\to0$, we get a strongly informative prior, which will override contributions made by finite data and therefore asserts $q_k=p_k$. 

For the case $p_k=\frac{1}{2}$ and $\theta=\frac{1}{2\alpha+1}\ge\frac{1}{3}$, we recover the above-mentioned non-informative priors: Laplace at $\theta=\frac{1}{3}$, Jeffreys at $\theta=\frac{1}{2}$ and in the extreme as $\theta\to1$, the Haldane prior, which gives maximum weight to the data. These effects will be shown below.

\subsection{Database}
\label{sec:db}
\def\Amat{\mathbf{A}}
We make provision in our calculation to optionally use a database of examples to help us infer values for $\qvec$. Let $\Amat=(\avec_1,\avec_2,\ldots,\avec_L)$ be a database of DNA profiles for $L$ different individuals, where the profile for individual $\ell$ is $\avec_\ell=(a_{1\ell},a_{2\ell},\ldots,a_{K\ell})$ and where $a_{k\ell}\in\{0,1\}$ is the binary state of locus $k$ of individual $\ell$. We assume 
the DNA profiles in $\Amat$:
\begin{itemize}
	\item have been sampled iid from the same population as the suspect and perpetrator and are therefore relevant to inferring the parameter $\qvec$, 
	\item but the individuals are distinct from the suspect and the perpetrator.
\end{itemize}

Our calculations will allow for the case of the empty database, where $L=0$.

\subsection{Likelihood}
Because of our independence assumptions in the model, the likelihood for $\qvec$, given the database $\Amat$ is:
\begin{align}
P(\Amat | \qvec) &= \prod_{\ell=1}^L \prod_{k=1}^K 
q_k^{a_{k\ell}}(1-q_k)^{1-a_{k\ell}} \\
&= \prod_{k=1}^K q_k^{n_k}(1-q_k)^{L-n_k}
\end{align}
where $n_k=\sum_{\ell=1}^L a_{k\ell}$ is the number of times locus $k$ has state 1 and $L-n_k$ is the number of times it has state 0.

\subsection{Posterior}
We can now infer the value of $\qvec$ by computing the posterior:
\begin{align}
P(\qvec|\Amat,\pivec) &= \frac
{P(\qvec|\pivec)P(\Amat|\qvec)}
{\int P(\qvec'|\pivec)P(\Amat|\qvec') \,d\qvec'} \\
&= \prod_{k=1}^K \frac
{ q_k^{\alpha_k+n_k-1}(1-q_k)^{\beta_k+L-n_k-1}}
{\int_0^1 q_k'^{\,\alpha_k+n_k-1}(1-q_k')^{\beta_k+L-n_k-1} \,dq_k'}\\
%{\prod_{k=1}^K \int P(q_k'|\alpha)P(\Amat|\qvec') \,dq_k'}} \\
%
\label{eq:qpost}
&= \prod_{k=1}^K \BD(q_k|\alpha_k+n_k,\beta_k+L-n_k)
\end{align}
where the integral in the denominator was solved by inspection, by recognizing the numerator as another beta distribution. This is due to the fact that the beta distribution is conjugate to the Bernoulli likelihood and therefore should result in a beta posterior. Notice that if the database is empty, then $n_k=L=0$ and the posterior is just the prior.

The prior parameters $\alpha_k$ and $\beta_k$ play the same roles mathematically as the event counts $n_k$ and $L-n_k$ and are consequently referred to as \emph{pseudo-counts}. The total pseudo count, $\alpha+\beta$ can be interpreted as the \emph{size} of some pseudo database, which is then effectively pooled with $\Amat$ by the additions in~\eqref{eq:qpost}.

In the alternative prior parametrization, $\frac{1-\theta}{\theta}p_k$ and $\frac{1-\theta}{\theta}(1-p_k)$ are the pseudo counts and $\frac{1-\theta}{\theta}$ is the size of the pseudo database.

The posterior $P(\qvec|\Amat,\pivec)$ represents our total state of knowledge about $\qvec$ and can be used in all calculations in place of the unknown $\qvec$.

\section{Forensic LR}
\def\rvec{\mathbf{r}}
\def\LR{\text{LR}}
We are given two DNA profiles: One for the \emph{suspect}, $\svec=(s_1,s_2,\ldots,s_K)$ and one for the \emph{perpetrator}, $\rvec=(r_1,r_2,\ldots,r_K)$. We work with two hypotheses and assume they are the \emph{only} possible explanations for the observed data $\svec,\rvec$:
\begin{itemize}
	\item The \emph{prosecution hypothesis}, $H_p$, asserts that suspect and perpetrator are the same person. 
	\item The \emph{defence hypothesis} $H_d$, asserts that they are different individuals. 
\end{itemize}
Below we compute the likelihoods under each hypothesis. For now, we assume that if they don't match, $\rvec\ne\svec$, then in the absence of DNA measurement errors, this proves deductively that $H_d$ is true and $H_p$ is false.

In the matched case, $\rvec=\svec$, however, we need probabilistic reasoning. The most natural way to do this would be to compute the \emph{posterior}, 
\begin{align}
    P(H_p|\rvec,\svec,\Amat,\pivec,\Pi) = 1-P(H_d|\rvec,\svec,\Amat,\pivec,\Pi)
\end{align}
where we have introduced the \emph{prior for the prosecution hypothesis}, 
\begin{align}
\Pi=P(H_p|\Pi)=1-P(H_d|\Pi)
\end{align} 
which is assigned by a reasoning process not involving DNA profiles. However, in the Bayesian paradigm for presenting evidence in court one equivalently considers the \emph{posterior odds for $H_p$ against $H_d$}, which can be separated\footnote{In the real world, this simple factorization applies only in a limited number of cases. If different alternative culprits, with different levels of relatedness to the suspect are considered, somewhat more general formulas have to be used, as explained in WEFDNA.} into two factors: \emph{likelihood ratio} and \emph{prior odds}, respectively representing the contributions of the DNA analysis and all other evidence not related to DNA:
\begin{align}
\label{eq:odds}
\frac{P(H_p|\rvec,\svec,\Amat,\pivec,\Pi)}{P(H_d|\rvec,\svec,\Amat,\pivec,\Pi)} &= \LR \, \frac{\Pi}{1-\Pi}
\end{align}
where 
\begin{align}
\LR &= \frac{P(\rvec,\svec|H_p,\Amat,\pivec)}{P(\rvec,\svec|H_d,\Amat,\pivec)}
\end{align} 
is referred to as \emph{the likelihood ratio}. It is then recommended that the end-goal of the forensic DNA analysis is to compute $\LR$, which can be done independently of $\Pi$. We derive expressions for both likelihoods below and then form the ratio.

Finally, notice that if $\LR=1$, then the DNA analysis is \emph{completely non-informative} about $H_p$ versus $H_d$: in this case the posterior (odds) is the same as the prior (odds).

\subsection{Prosecution likelihood}
Under the prosecution hypothesis, $\rvec$ and $\svec$ come from the same individual, so that $P(\rvec,\svec|H_p,\qvec)=\delta(\rvec,\svec)P(\svec|\qvec)$, where $\delta(\rvec,\rvec)=1$, or $\delta(\rvec,\svec)=0$ if $\rvec\ne\svec$. Since we are not given $\qvec$, but instead we are given the prior $\pivec$ and the database $\Amat$, we must condition on what we have and instead compute:
\begin{align}
P(\rvec,\svec|H_p,\pivec,\Amat)
&=\delta(\rvec,\svec) P(\svec|\pivec,\Amat)
\end{align}
where 
\begin{align}
P(\svec|\pivec,\Amat) &=
\int_0^1\int_0^1\cdots\int_0^1 
P(\svec|\qvec)P(\qvec | \pivec,\Amat)\,dq_1 dq_2\cdots dq_K \\
&= \int_Q P(\svec|\qvec)P(\qvec | \pivec,\Amat)\,d\qvec
\end{align} 
where $Q$ is short-hand for the $K$-cube over which we are integrating. Note $P(\svec|\pivec,\Amat)$ is called the \emph{predictive distribution} for $\svec$, because it predicts the value of an as yet unseen profile, given that we have already seen the profiles in $\Amat$. Again by virtue of the conjugate prior, the predictive distribution can be found in closed form:
\begin{align}
P(\svec|\pivec,\Amat) &= \int_Q P(\svec |\qvec) P(\qvec | \pivec,\Amat) \,d\qvec \\
&= \prod_{k=1}^K \int_0^1 q_k^{s_k}(1-q_k)^{1-s_k} \frac{q_k^{\alpha_k+n_k-1}(1-q_k)^{\beta_k+L-n_k-1}}{B(\alpha_k+n_k,\beta_k+L-n_k)} \,dq_k  \\
&= \prod_{k=1}^K \frac%
{\int_0^1 q_k^{\alpha_k+s_k+n_k-1}(1-q_k)^{\beta_k+L+1-s_k-n_k-1} \,dq_k}
{B(\alpha_k+n_k,\beta_k+L-n_k)} \\
&= \prod_{k=1}^K \frac
{B(\alpha_k+s_k+n_k,\beta_k+L+1-s_k-n_k)}
{B(\alpha_k+n_k,\beta_k+L-n_k)} \\
&= \prod_{k=1}^K P(s_k|\pi_k,n_k,L)
\end{align}
Now we can expand the beta functions in terms of gamma functions and simplify the ratios of gammas with the identity $\Gamma(x+1)=x\Gamma(x)$, to find the predictive probability:\footnote{Notice~\eqref{eq:pred1i} agrees with equation 5.6 on page 64 in WEFDNA.} 
\begin{align}
\label{eq:pred1n}
P(s_k=1|\pi_k,n_k,L)
&= \frac{\alpha_k+n_k}{\alpha_k+\beta_k+L} \\
\label{eq:pred1i}
&= \frac{(1-\theta)p_k+\theta n_k}{(1-\theta)+\theta L} 
\end{align}
For the informative prior case, notice that $\theta$ gives \emph{interpolation} weights between data and the prior parameter $p_k$. At the one extreme if $\theta\to1$ (Haldane prior), we disregard the prior parameter $p_k$ and end up with just the data proportion $\frac{n_k}{L}$. At the other extreme if $\theta=0$, we disregard the data $\Amat$ and end up with the prior parameter $p_k$. (If we use the non-informative Laplace prior, with $\alpha_k=\beta_k=1$, then~\eqref{eq:pred1n} is known as Laplace's \emph{rule of succession}.) Finally, the predictive probability\footnote{Notice $P(s_k=0|\alpha_k,\beta_k,n_k,L)+P(s_k=1|\alpha_k,\beta_k,n_k,L)=1$.} for the event $s_k=0$ is:
\begin{align}
P(s_k=0|\pi_k,n_k,L)
&= \frac{\beta_k+L-n_k}{\alpha_k+\beta_k+L} \\
&= \frac{(1-\theta)(1-p_k)+\theta (L-n_k)}{(1-\theta)+\theta L} 
\end{align}
Note that even for an empty database ($L=n_k=0$), our assumption $\alpha_k,\beta_k>0$ guarantees non-zero predictive probabilities.

\subsection{Defence likelihood}
Under the defence hypothesis, $\rvec$ and $\svec$ come from different individuals and their probabilities are independent given $\qvec$, so that $P(\rvec,\svec|\qvec)=P(\rvec|\qvec)P(\svec|\qvec)$. However, $\qvec$ is not given, so the independence no longer holds: knowledge of one profile changes the probability for $\qvec$, which in turn changes the probability for the other profile. This dependency is automatically taken care of by applying the rules of probability theory by integrating out the unknown $\qvec$:
\begin{align}
P(\rvec,\svec|H_d,\pivec,\Amat) &= 
\int_Q P(\rvec |\qvec) P(\svec |\qvec) P(\qvec | \pivec,\Amat) \,d\qvec \\
&= \prod_{k=1}^K \frac
{B(\alpha_k+s_k+r_k+n_k,\beta_k+L+2-r_k-s_k-n_k)}
{B(\alpha_k+n_k,\beta_k+L-n_k)} \\
&= \prod_{k=1}^K P(r_k,s_k|\pi_k,n_k,L)
\end{align}
where we can expand and simplify again to find the predictive probability: 
\begin{align}
\label{eq:pred11n}
P(r_k=s_k=1|\pi_k,n_k,L) &= \frac{\alpha_k+n_k}{\alpha_k+\beta_k+L} \; \frac{\alpha_k+n_k+1}{\alpha_k+\beta_k+L+1} \\
&=P(r_k=1|\pi_k,n_k,L) P(s_k=1|\pi_k,n_k+1,L+1) 
\end{align}
Notice the similarity between the two factors in the RHS: the right factor is obtained from the left by adding 1's to the observation counts. Notice also that if $\alpha+n_k\gg1$, then $P(r_k=s_k=1|\pi_k,n_k,L)\approx P(r_k=1|\pi_k,n_k,L)P(s_k=1|\pi_k,n_k,L)$, making the two events \emph{almost} independent. 

The probability for the other event of interest\footnote{We don't need the events $(0,1)$ and $(1,0)$ here, because we are interested in the case where profiles match.} is obtained similarly as:
\begin{align}
P(r_k=s_k=0|\pi_k,n_k,L) &=
P(r_k=0|\pi_k,n_k,L) P(s_k=0|\pi_k,n_k,L+1)
\end{align}

\subsection{LR}
Forming the likelihood-ratio, we find:
\begin{align}
\LR&=\frac{P(\rvec,\svec|H_p,\pivec,\Amat)}
{P(\rvec,\svec|H_d,\pivec,\Amat)}  
= \prod_{k=1}^K \LR_k(r_k,s_k)
\end{align}
where
\begin{align}
\LR_k(r,s)&= \frac
{\delta(r,s)P(s|\pi_k,n_k,L)}
{P(r,s|\pi_k,n_k,L)} \\
&= \frac{\delta(r,s)P(s|\pi_k,n_k,L)}
{P(s,s|\pi_k,n_k,L)} \\
&=\frac{\delta(r,s)P(s|\pi_k,n_k,L)}{P(s|\pi_k,n_k,L)P(s|\pi_k,n_k+s,L+1)} \\
&=\frac{\delta(r,s)}{P(s|\pi_k,n_k+s,L+1)}
\end{align}
More explicitly, for the mismatched cases we have
\begin{align}
\LR_k(0,1)=\LR_k(1,0)=0
\end{align}
and for the matched cases we have
\begin{align}
\LR_k(1,1) &=\frac{\alpha_k+\beta_k+L+1}{\alpha_k+n_k+1}\,, &
\LR_k(0,0) &= \frac{\alpha_k+\beta_k+L+1}{\beta_k+L+1-n_k}
\end{align}
or, with the other prior parametrization:
\begin{align}
\label{eq:lr11}
\LR_k(1,1) &= \frac{(1-\theta)+\theta(L+1)}{(1-\theta)p_k+\theta(n_k+1)} \\
\intertext{and}
\label{eq:lr00}
\LR_k(0,0) &= \frac{(1-\theta)+\theta(L+1)}{(1-\theta)(1-p_k)+\theta(L+1-n_k)}
\end{align}
Notice again, that $\theta$ interpolates between data and the prior parameter $p_k$. The minimum value (for the matched case $r_k=s_k$) is 1. This is a consequence of the error-free measurement assumption. If non-zero error probabilities were considered, values of less than 1 would be possible.

\section{Plug-in recipe}
In this section, we shall refer to:
\begin{itemize}
	\item One or more \emph{reference populations}, from which one or more databases are drawn to help to estimate the parameters $p_k$ and $\theta$ for an \emph{informative} prior.
	\item The \emph{relevant population}, from which the suspect and perpetrator were drawn. 
\end{itemize}
In the general case, all these populations are assumed different from each other in the sense that locus state \emph{frequencies may differ} between them.

For forensic DNA applications, WEFDNA motivates a \emph{plug-in} recipe to compute the $\LR$, where values for $\theta$ and the $p_k$ are point-estimates made from one or more reference databases. In this recipe, the $p_k$ are representative of the frequencies in the \emph{reference} populations, while the value of $\theta$ is chosen to reflect by how much the corresponding frequencies in the \emph{relevant} population may differ. Small values of $\theta$ encode small expected differences and larger values encode larger expected differences. WEFDNA motivates for values in the range $1\%\le\theta\le5\%$ to be used for most applications.

Our database $\Amat$, as defined in section~\ref{sec:db}, is assumed to be drawn from the \emph{relevant} population, but in the usual forensic scenario, additional profiles from the relevant population are not available. In our notation, this means $\Amat$ is empty. 

In summary, in the WEFDNA plug-in recipe we set $L=n_k=0$, the $p_k$ are generally different from $\frac{1}{2}$ and $\theta$ is smallish. This forms an \emph{informative} prior for the $q_k$. This gives, for $r_k=s_k$:
\begin{align}
\LR_k(1,1) &= \frac{1}{(1-\theta)p_k+\theta}\,&
\LR_k(0,0) &= \frac{1}{(1-\theta)(1-p_k)+\theta}
\end{align}

\section{Fully Bayesian recipe}
Now we turn to the main purpose of this document, namely to explore a fully Bayesian recipe, where we start with a \emph{non-informative} prior and use only the given data, $\Amat,\rvec,\svec$, to infer the model parameter.

It must be emphasized that this fully Bayesian recipe cannot be used as is to replace the plug-in recipe, because here we use the luxury of database $\Amat$, sampled from the relevant population. As noted above, in a realistic scenario, we do not have this luxury: instead we have to make do with data sampled from some other, somewhat different, reference population. Although a fully Bayesian recipe could in principle be derived for this more realistic scenario, this would come at the cost of a considerable increase in both conceptual difficulties as well as computational complexity.

In this section, therefore we assume we \emph{do} have a database, $\Amat$, sampled from the relevant database and the only difficulty that remains is to choose the non-informative prior.

\subsection{Which prior?}
We are now faced with making a choice amongst the different flavours of non-informative priors. That is, we have to choose $\alpha_k$ and $\beta_k$, or equivalently $p_k$ and $\theta$.

We concede that we are choosing a prior under the perhaps arbitrary constraint that it should be a beta distribution. A more thorough motivation for the prior should perhaps involve solving functional equations in the style of PTLOS. We feel however that the beta distribution already provides a rich enough space for the choice of prior. Moreover, as mentioned above, the non-informative Haldane, Jeffreys and Laplace priors all members of the beta family.

To start, we motivate the choice $\alpha=\alpha_k=\beta_k$, or equivalently $p_k=\frac{\alpha_k}{\alpha_k+\beta_k}=\frac{1}{2}$. Before we have seen any data, all loci are on an equal footing, so that the priors for all $k$ must be the same. Next consider a database $\Amat$ with an equal number of 0's and 1's for some locus $k$, so that $n_k=L-n_k$. In this situation, there is no reason to prefer one state to the other, so that the model parameter posterior should satisfy the symmetry condition: $P(q_k|\alpha_k,\beta_k,L,n_k) = P(1-q_k|\alpha_k,\beta_k,L,n_k)$, which is obtained at $\alpha_k=\beta_k$. Another way to see this is simply to require $\LR_k(0,0)=\LR_k(1,1)$ when $n_k=L-n_k$.

Now we have $p_k=\frac{1}{2}$ and we still need to choose $\theta$. To do this, consider the case of the \emph{empty} database, with $L=n_k=0$, for which case we still want our recipe to give a sensible answer. Now~\eqref{eq:lr11} and~\eqref{eq:lr00} give: 
\begin{align}
\LR_k(1,1)&=\LR_k(0,0)=\frac{1}{(1-\theta)\frac{1}{2}+\theta}=\frac{2}{1+\theta}
\end{align}
When $\Amat$ is empty, we now argue that we don't even know whether the locus state varies in the population. So we are not justified in concluding that the match at the locus modifies the probabilities for $H_p$ vs $H_d$. If we maximize $\theta$ at the limit $\theta\to1$, then we obtain the non-informative value of $\LR_k=1$, so that the DNA evidence is effectively \emph{disregarded}.

\subsection{Analysis}
Here we analyse the behaviour of $\LR_k(r_k,s_k)$, when $r_k=s_k$ and $\theta=1$. We get:
\begin{align}
\LR_k(1,1) &= \frac{L+1}{n_k+1}, &
\LR_k(0,0) &= \frac{L+1}{L+1-n_k}
\end{align}
We make several observations:
\begin{itemize}
  \item The matched likelihood ratios are bounded: $1\le LR_k(s,s) \le L+1$. We have already commented on the lower bound. The upper bound is determined by the database size, $L$. This makes intuitive sense, the larger the database, the more our maximum confidence grows. Note however, that this maximum should be a relatively rare occurrence, as shown below.
	\item For an empty database, if $L=n_k=0$, then as discussed, $\LR_k(1,1)=\LR_k(0,0)=1$.
	\item For a non-empty database, as long as a locus $k$ has the same state in all of the observed data, $\Amat,\rvec,\svec$, then the $\LR$ is \emph{still} unity: If $n_k=L$, then $\LR_k(1,1)=1$ and if $n_k=0$, then $\LR_k(0,0)=1$.
	\item Conversely, for a given database size $L$, the maximum $\LR$ value is reached when the locus state observed in $s_k=r_k$ has never been observed in $\Amat$. This implies the trait shared by the suspect and perpetrator is \emph{rare}. The larger the database size, $L$, the more we are convinced of the rarity and the more we are convinced of the identity of suspect and perpetrator.
\item For a large database, where both $n_k\gg1$ and $L-n_k\gg1$, the likelihood ratio for $s_k=r_k$ is the inverse of the frequency of the corresponding event in the database: $\LR_k(1,1)\approx\frac{L}{n_k}$ and $\LR_k(0,0)\approx\frac{L}{L-n_k}$. 
\end{itemize}
We can briefly compare this recipe to a very naive recipe, where we simply assign $q_k=\frac{n_k}{L}$, irrespective of the size of the database. This would give $\LR(1,1)=\frac{L}{n_k}$ and $\LR_k(0,0)=\frac{L}{L-n_k}$. This agrees with the last case above of the Bayesian recipe, but in any other cases it could give overconfident results. In particular, if $n_k=0$, or $n_k=L$, one could get infinite $\LR$ values, which would be ridiculous in the extreme if $L=1$. The fully Bayesian recipe agrees with the naive recipe when data is plentiful, but continues to give sensible answers even when the data gets scarce to the point of vanishing.  

\subsubsection{Comment on Haldane prior}
With a more realistic DNA model, where each STR locus has two independent sides (paternal and maternal), we can gain some extra insight into the nature of the Haldane prior. In this case, it can be shown (WEFDNA, section 6.2.2) that when $L=0$, the LR for a locus can nevertheless reach a maximum of 3. If the paternal and maternal sides are the same, then we get LR=1, but if they are different, we get LR=3. From this fact and the third bullet above, we learn that: 
\begin{quote}
The LR at locus $k$ becomes non-informative ($\LR_k=1$) under the Haldane prior, if and only if no state change has been observed at locus $k$ in \emph{all} of the data, $\Amat,\rvec,\svec$. 
\end{quote}
One may argue that loci used for forensic DNA profiling have been chosen for the purpose of giving good discrimination between individuals, precisely because they \emph{do} vary appreciably between individuals and that therefore the Haldane prior is too extreme. However, we are concerned here with \emph{sub-populations}, about which we cannot assume that every locus is informative---it may well be that a certain locus is constant over the whole sub-population. We therefore argue that the behaviour of the Haldane prior is appropriate: the LR for a locus remains non-informative ($\LR_k=1$), until we have observed at least one state change in our data. 

\end{document}